\documentclass[12pt]{article}
\begin{document}
\title{Relativistic correction to the Bohr radius and electron distance expectation value via Dirac Equation}
\maketitle
\author{J. Buitrago}
\bigskip

\noindent{University of La Laguna, Faculty of Physics. 38205, La Laguna (Tenerife) Spain. jgb@iac.es, jbuitrag@ull.es}

\begin{abstract}
%{\bf Abstract.}
In this article and beginning with the Dirac solution to the Hydrogen atom in its ground state, the exact results corresponding to the expectation value of the distance of the electron to the proton and the maximum probability distance are found. For hydrogen-like atoms and in contrast to the non relativistic Bohr radius and expectation values, for $Z=\frac{\sqrt3}{2}\frac{1}{\alpha}$ (close to 118), the expectation distance to the nucleus is zero thus putting an end to the periodic table for Z = 118 nicely matching the last element discovered (Oganesson).
\end{abstract}
\noindent{{\bf keywords:} Dirac Equation, relativistic quantum mechanics, Bohr radius}

\bigskip
\section{Introduction}

From its appearance in 1927, the Dirac Equation has been one of the more, if not the most, studied equation in modern physics. One of the more illuminating aspects is the treatment of the Hydrogen Atom which gives quite a few relativistic features not appearing in the paradigmatic study found in NR quantum mechanics. It comes as a surprise that until now and as far as I know, nobody has cared to obtain such a fundamental expression as the relativistic analogue of the Bohr radius not even the expectation value for the distance to the nucleus in the ground state. As both results (to my knowledge never mentioned in any graduate text book) are of conceptual, pedagogical  and practical value, specially for the students that find in this case perhaps the only one in which former fundamental ideas such as the probability interpretation of the wave function can be carried over to the relativistic arena. Analogies with the NR treatment go even to the curious case in which, as in the old  Bohr study of the hydrogen atom, the exact result for the binding energy of the ground state can be found by the same kind of simple minded approach found in the Bohr classical study (see [1]). It is the subject of this short communication to fill these curious gaps and obtain analytically  both the expectation value and the relativistic expression of the Bohr Radius thus alleviating the new comer student the feeling of having to deal with an esoteric abstruse theory with effects such as positive and negative energy states or the puzzling Zitterbewegung [3] (to number only a few). Indeed I consider that what the reader can be found below should be included in future text books dealing with the Dirac Equation.

\section{Relativistic Bohr radius and distance expectation values of the hydrogen atom}

In natural units ($\hbar=c=1$), the normalized spin-up wave function solution of the Dirac equation for the Hydrogen atom in its ground state is [2]

\begin{equation}\label{sol1}
\psi(r,\theta,\varphi)=\frac{(2m\alpha)^{3/2}}{\sqrt {4\pi}}\sqrt\frac{1+\gamma}{2\Gamma(1+2\gamma)}(2m\alpha r)^{\gamma-1}e^{-m\alpha r}\left(\begin{array}{c}
1 \\
0 \\
\frac{i(1-\gamma)}{\alpha}\cos\theta \\
\frac{i(1-\gamma}{\alpha}\sin \theta^{i\varphi} \\

\end{array}\right),
\end{equation}
with
$$
\gamma = \sqrt {1-\alpha^2}.
$$
$\Gamma$ is the Gamma function of argument $(1+2\gamma)$, $m$ the mass of the electron and $\alpha$ the fine structure constant.
From (1), and its complex conjugate, it is straightforward to see that the distribution of probability is isotropic:
\begin{equation}
	\psi^*\psi \propto \left [1+\frac{(1-\gamma)^2}{\alpha^2}\cos^2\theta+\frac{(1-\gamma)^2}{\alpha^2}\sin^2\theta \right].
\end{equation}

The next step is to find the distance to the nucleus where the probability of finding the electron is maximal and the expectation value. 
From the previous expression and (1):
\begin{equation}
	\psi^*\psi= \frac{1}{4\pi}8m^3\alpha^3\frac{1}{\Gamma(1+2\gamma)}(2m\alpha r) e^{-2m\alpha r}.
\end{equation}
The expectation value
\begin{equation}
	<r>=\int_0^\infty \psi^* r \psi 4\pi r^2 dr,
\end{equation}
follows by making $2m\alpha r=x$ and solving the integral
\begin{equation}
	\int_0^\infty e^{-x} x^{2\gamma +1}dx=\Gamma(2\gamma+2),
\end{equation}
together with the recursion formula
$$
\Gamma (2\gamma+2)=(2\gamma+1)\Gamma(2\gamma+1).
$$

 The calculations can be carried out without much difficulty with the result:

\begin{equation}
	<\psi|r|\psi>= \frac{1}{m}\left[\sqrt{\frac{1}{\alpha^2}-1}+ \frac{1}{2\alpha}\right ].
\end{equation}
In order to find the radial distance of the electron to the nucleus where the probability of finding the electron is maximum, we consider first:

\begin{equation}
	\psi^*\psi r^2 =
\frac {2m^{3}\alpha ^{3}}{\pi }\frac {\left( 2m\alpha \right) ^{2\left( \gamma -1\right) }}{\Gamma \left( 1+2\gamma \right) } r ^{2\gamma }e^{-2m\alpha r}.
\end{equation}
As $\Gamma(1+2\gamma)$ is a magnitude depending only on $\gamma$, the radius of maximum probability comes out after solving for $r$ the equation resulting from:
$$
\frac{d}{dr}[r^{2\gamma}.e^{-2m\alpha r}] = 0.
$$
The result is
\begin{equation}
r_{max}= \frac{1}{m}\sqrt{\frac{1}{\alpha^2}-1}.
\end{equation}
In the non relativistic approximation equations (6) and (8) reproduce the familiar results of NR quantum mechanics.
\bigskip

The last expressions are also valid for Hydrogen-like atoms making the substitution: 
$$
\alpha \longrightarrow Z\alpha.
$$
\bigskip
For $Z=1/ \alpha$, which is about 137, $r_{max} =0$.
\bigskip

If the former estimation seems somewhat crude it is possible to obtain a better sensible result. With the same substitution for $\alpha$ and equating to zero the right hand side of (6) for the expectation value, we find solving for $Z$:
$$
Z=\frac{\sqrt3}{2}\frac{1}{\alpha},
$$
which is close to 118.

The last, officially recognized in 2016, element is Oganesson (Og) with Z=118.
\bigskip

{\noindent \bf Bibliography}
\bigskip

\noindent{
[1] J. Buitrago, 1985, "About a heuristic solution to the binding  energy of the hydrogen atom". European Journal of Physics Vol.6 n 3, 209}

\bigskip

{\noindent [2] J.D. Bjorken and S.D. Drell "Relativistic quantum mechanics". 1964 McGraw-Hill}

\bigskip
{\noindent [3] For a new interpretation of this phenomenon see: J. Buitrago, 2018 "Electron internal energy and motion as consequence of local U(1) gauge invariance". Results in Physics (Elsevier) 11, 138-143}
	
\end{document}